\documentclass[9pt]{article}
\usepackage{spconf,amsmath,graphicx, cite}

\usepackage{amsfonts}
\usepackage{url}
\usepackage{color}
\usepackage[english]{babel}
\usepackage[T1]{fontenc}
\usepackage[utf8]{inputenc}
\usepackage{booktabs} 
\usepackage{algorithm}
\usepackage{algorithmic} 
\usepackage{nicefrac}
\usepackage{overpic}
\usepackage{tikz}
\usepackage{multirow}
 \usepackage{diagbox}

\DeclareRobustCommand\onedot{\futurelet\@let@token\@onedot}
\def\@onedot{\ifx\@let@token.\else.\null\fi\xspace}

\newcommand{\ourname}{ByteCover3}

\newcommand\blfootnote[1]{%
  \begingroup
  \renewcommand\thefootnote{}\footnote{#1}%
  \addtocounter{footnote}{-1}%
  \endgroup
}

\newcommand\specparen[2]{%
  \def\Krn{\kern1ex}%
  \def\useanchorwidth{T}%
  \setbox0=\hbox{\Krn\stackengine{0pt}{\scriptstyle#1}{\scriptstyle#2}{O}{c}{F}{F}{S}}%
  \stackon[2pt]{\stackunder[2pt]{)}{\makebox[\wd0][l]{\Krn$\scriptstyle#1$}}}%
                                   {\makebox[\wd0][l]{\Krn$\scriptstyle#2$}}%
}

\title{ByteCover3: Accurate Cover Song Identification on Short Queries}
\name{Xingjian Du$^1$, Zijie Wang$^{2\dagger}$, Xia Liang$^{1\dagger}$,  Huidong Liang$^3$, Bilei Zhu$^1$, Zejun Ma$^1$}
\address{$^1$ByteDance\quad $^2$Zhejiang University\quad$^3$University of Oxford}
\begin{document}
%
\maketitle

\begin{abstract}
Deep learning based methods have become a paradigm for cover song identification (CSI) in recent years, where the ByteCover systems have achieved state-of-the-art results on all the mainstream datasets of CSI. However, with the burgeon of short videos, many real-world applications require matching short music excerpts to full-length music tracks in the database, which is still under-explored and waiting for an industrial-level solution. In this paper, we upgrade the previous ByteCover systems to \emph{\ourname{}} that utilizes local features to further improve the identification performance of short music queries. \ourname{} is designed with a local alignment loss (LAL) module and a two-stage feature retrieval pipeline, allowing the system to perform CSI in a more precise and efficient way. We evaluated \ourname{} on multiple datasets with different benchmark settings, where \ourname{} beat all the compared methods including its previous versions. 
\blfootnote{$\dagger$ These authors contributed equally.}
\end{abstract}


\begin{figure*}
	\centering
	\vspace{-2em}
	\includegraphics[width=0.9\textwidth]{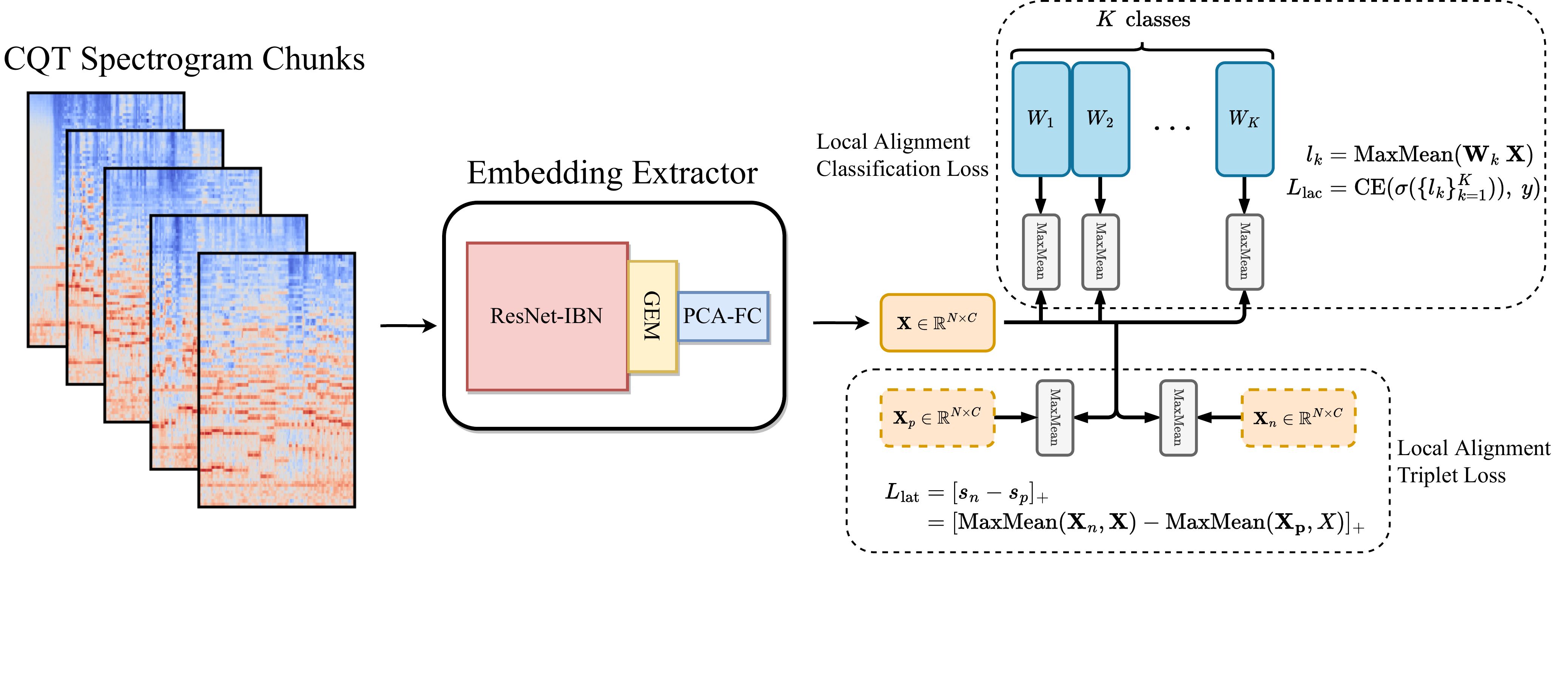}
	\vspace{-4em}
        \caption{The architecture of \ourname{}. \ourname{} uses an embedding extractor to extract local features from the CQT chunks and optimizes the model using the LAL loss. The LAL loss consists of a LAL classification loss and a LAL triplet loss.}
	\vspace{-1em}
	\label{fig:overall}

\end{figure*}

\section{Introduction}

Recent years have witnessed a successful use of deep learning methods in the task of cover song identification (CSI), i.e., finding cover versions of a given music track in a music database. These methods generally formulate CSI as either a classification problem~\cite{yu2019temporal, yu2020learning, xu2018key} or a metric learning problem~\cite{yesiler2020accurate}, or a combination of both~\cite{du2021bytecover,du2022bytecover2,hu2022wideresnet}, and then train deep neural networks to learn low-dimensional features from different representations of audio. The features are then indexed and retrieved, where the distances between features are used to measure the similarities of songs. Deep learning models have proved their capability in learning discriminative and robust features, boosting the accuracy of CSI by a large margin compared to traditional methods based on handcrafted features.



Despite of the promising progress above, one challenge remains for CSI in real-world applications, which is that most existing methods only consider situations when the query and the database item are both full-length music tracks with a typical duration of several minutes. Nevertheless, in many real-world scenarios, the task is to identify short music queries which are, for example, tens of seconds long, against full-length database songs. For instance, massive short videos with less-than-one-minute length have been created and uploaded to short video platforms such as TikTok in the past years, where a large proportion of these videos are accompanied by a carefully-selected music track that may be a remixed or cover fragment of an original music. For copyright management and reporting purpose, platforms need to identify these cover fragments. Unfortunately, as shown in section~\ref{sec:exp}, most existing CSI systems failed in the experiments of identifying short queries.  
An explanation for such incapability is that current CSI methods are mostly equipped with a global pooling layer to aggregate the information from all time sections, and then generate a global embedding for each song as the CSI feature. However, when matching a short music clip to a full-length recording, there will be some irrelevant sections in the full-length recording that may create noises in the embedding, which pose a negative impact on the similarity measurement between features. \par
To solve this problem, an intuitive idea is to extract local features from both queries and database items, and calculate the similarity between two songs based only on the local features that have high matching score, thus avoiding the interference from irrelevant sections. This idea has been considered in a few traditional CSI works~\cite{muller2005audio, casey2008analysis, grosche2012toward, rafii2014audio, cai2016two}. However, due to the limited discriminative capacity and robustness of handcrafted features, as well as the high complexity in performing efficient indexing and retrieval, traditional CSI methods are generally only applicable for small databases (e.g., the database used in \cite{grosche2012toward} contains only 2,484 recordings) or simplified scenarios (e.g., \cite{rafii2014audio} can only identify live versions of songs from known artists), while their performances for the general CSI task against large databases are poor (or have not been reported). In \cite{zalkow2021efficient}, Zalkow \emph{et al.} made the first attempt to employ deep learning for short-query CSI and used convolutional neural networks (CNNs) to compress the input features. However, the databases used in \cite{zalkow2021efficient} are still limited to thousands of audio recordings or even less, and its usability in real-world applications is unverified.


In this paper, we extend our previous works of ByteCover~\cite{du2021bytecover} and ByteCover2~\cite{du2022bytecover2}, and present the new version of our CSI system, \emph{ByteCover3}, to solve the problem of identifying short music queries against a industry-scale database of full-length recordings. Different from existing works that rely on global features~\cite{yesiler2020accurate,yu2020learning,doras2019cover,du2021bytecover,du2022bytecover2}, ByteCover3 is designed to learn a set of deep local embeddings (or features) from each audio and uses the matching of local embeddings to accomplish the identification of short queries against full songs. To optimize the matching of local features, we propose a new loss termed \emph{local alignment loss} (LAL) and apply it in our multi-loss paradigm first introduced in ByteCover \cite{du2021bytecover}. LAL constitutes one of the major contributions of ByteCover3, and by using the LAL loss, the performance of CSI can be significantly improved. Moreover, to improve the efficiency of feature matching, a two-stage feature retrieval pipeline, which consists of an approximate nearest neighbor (ANN) stage and a re-ranking stage, is also designed. This consists of our second contribution.


\vspace{-0.2cm}
\section{ByteCover3}
\label{sec:approach}
\vspace{-0.8em}
The overall architecture of \ourname{} is illustrated in Fig. \ref{fig:overall}. \ourname{} is inherited from ByteCover and ByteCover2 and adopts a multi-loss learning paradigm for CSI. One of the major settings that differs it from previous works lies in the use of local features. In this section, we first describe the extraction of local features and then introduce our main contributions, i.e., the LAL loss for local feature matching and the two-stage feature retrieval pipeline. 
\vspace{-1em}
\subsection{Local Feature Extraction}
To extract local features from each recording, we first resample the audio to $22,050$ Hz and split it into $N$ short chunks with a length of $20$ seconds and a overlap of $10$ seconds. For each chunk, we calculate a constant-Q transform (CQT) spectrogram with the number of bins per octave and the hop size set to $12$ and $512$ respectively, using Hann window as the window function. The CQT spectrograms are then downsampled with an averaging factor of $100$ along the time axis to reduce the computation cost. Therefore, the input audio is processed into a compressed $N$-chunk CQTs $\mathbf{S} \in \mathbb{R}^{N \times F \times T}$, where $N$ is the number of chunks, $F$ is the number of CQT bins ($84$ in our setting), and $T$ is the number of frames in each chunk.

The ResNet-IBN model \cite{du2021bytecover}, which replaces the residual connection blocks of ResNet50~\cite{he2016deep} with the instance-batch normalization (IBN) blocks, is then applied as the backbone to extract local embeddings from the input CQTs. In ByteCover3, our ResNet-IBN follows the original ByteCover setting, except that a 3-D input $\mathbf{S} \in \mathbb{R}^{N \times F \times T}$ is taken instead of the original 2-D input. The output of ResNet-IBN before the global generalized mean (GeM) pooling layer is hence a 4-D embedding ${\mathbf{Z}} \in \mathbb{R}^{N \times C \times H \times W}$, where $C$ is the number of output channels, $H$ and $W$ are the spatial sizes along the frequency and time axes respectively. In practice, we set $C = 2048$, $H = 6$ and $W=\nicefrac{T}{8}$. Finally, the temporal and frequency axes of $\mathbf{Z}$ are integrated by the GeM pooling operation and a dimensionality reduction module, i.e., PCA-FC \cite{du2022bytecover2}, is utilized on the channel dimension to obtain the compacted final embedding ${\mathbf{X}} \in \mathbb{R}^{N\times 512}$, which contains $N$ local features from the original audio, as opposed to ByteCover and ByteCover2 that only adopted a single global embedding.
\vspace{-1em}
\subsection{The Local Alignment Loss}
Existing CSI methods generally employ either a classification loss (e.g., softmax loss) or a metric learning loss (e.g., triplet loss) or a combination of them as the optimization objective during training. Nevertheless, as these methods only rely on a single global vector for each music track, their loss functions are limited to measuring similarities between two vectors (e.g. dot product or cosine similarity). Whereas in our case, we wish to compare two sequences of vectors that contain different number of local features, which requires a new similarity measure. To address this problem, we propose a novel loss design called Local Alignment Loss (LAL) $\mathcal{L}_{\operatorname{lal}}$ consisting of a classification loss $\mathcal{L}_{\operatorname{lac}}$ and a triplet loss $\mathcal{L}_{\operatorname{lat}}$.


We first introduce a similarity measure termed \emph{MaxMean} inspired by~\cite{cai2016two}: let $\mathbf{X} \in \mathbb{R}^{M\times C}$ and $\mathbf{Y} \in \mathbb{R}^{N\times C}$ denote two $C$-dimensional feature sequences that each contain $M$ and $N$ local features ($M$ and $N$ could be highly different). For each local feature $\{\mathbf{x}_i\}_{i=1}^M \in \mathbb{R}^{1\times C}$ in $\mathbf{X}$, we calculate the cosine similarity between $\mathbf{x}_i$ and all the local features $\{\mathbf{y}_i\}_{j=1}^N \in \mathbb{R}^{1\times C}$ in $\mathbf{Y}$, and regard the maximal value as the similarity measure $s_i$ for $\mathbf{x}_i$:
\begin{equation}
s_i = \max(\cos(\mathbf{x}_i, \mathbf{y}_j)), ~j=1,\dots,N,
\end{equation}
and the final similarity score is obtained by taking average over all the similarity measures, i.e., $\operatorname{MaxMean}(\mathbf{X}, \mathbf{Y})=\frac{1}{M}\sum_{i=1}^M s_i$. The shorter local feature is always regraded as the first operand, because the \textit{MaxMean} operator is non-commutative. Since only the maximal value of all the matching scores from $\mathbf{x}_i$ to $\mathbf{Y}$ is considered, we can avoid the distractions of local features in $\mathbf{Y}$ that are irrelevant to $\mathbf{x}_i$. 

With the \emph{MaxMean} measure described above, we then illustrate how the original classification loss in previous ByteCover~\cite{du2021bytecover} is transformed to the novel LAL. Recall in ByteCover, the classification loss $\mathcal{L}_{cls}$ is defined as:
\begin{align}
\mathcal{L}_{cls} = \operatorname{CE}(\sigma(\mathbf{W}\mathbf{f}^\mathrm{T}),~ y) = \operatorname{CE}(\sigma(\{\mathbf{w}_k\mathbf{f}^\mathrm{T}\}_{k=1}^{K}),~ y),
\end{align}
where $\operatorname{CE}(\cdot,\cdot)$ is the cross entropy and $\sigma(\cdot)$ is the softmax function. We denote $y$ as the ground-truth label, $\mathbf{f} \in \mathbb{R}^{1\times C}$ as the global feature extracted from ResNet-IBN, and $\mathbf{W}\in\mathbb{R}^{K\times C}$ as the weight matrix in the linear layer before softmax that contains $K$ weight vectors $\{{\bf w}_k\}_{k=1}^K$ for classification.

To adapt $\mathcal{L}_{cls}$ to the novel \emph{MaxMean} measure with the local features, we draw inspiration from \cite{sun2020circle} and consider $\mathbf{w}_k$ as a proxy feature representation of the $k^\mathrm{th}$ class. In this sense, the result of $\mathbf{w}_k\mathbf{f}^\mathrm{T}$ can be interpreted as the similarity score between two features $\mathbf{w}_k$ and $\mathbf{f}$ based on dot product, which we argue can be replaced by the \emph{MaxMean} metric. Specifically, our new local alignment classification loss $\mathcal{L}_{lac}$ is written as:
\begin{align}
\mathcal{L}_{lac}  & = \operatorname{CE}(\sigma(\{\operatorname{logit}_k\}_{k=1}^K), y),\\
\operatorname{logit}_k  & = \operatorname{MaxMean}(\mathbf{X}, \mathbf{W}_k),
\end{align}
where ${\bf X} \in \mathbb{R}^{N \times C}$ is the final embedding with $N$ local features extracted by ResNet-IBN, $\mathbf{W} \in \mathbb{R}^{K \times L \times C}$ is a trainable weight matrix in the linear layer before softmax and ${\bf W}_k \in \mathbb{R}^{L \times C}$ denotes the proxy representation for class $k$.

In addition to the classification loss, a triplet loss was also used in ByteCover, which is simply modified in ByteCover3 by replacing the Euclidean distance with \emph{MaxMean} metric:
\begin{equation}
\mathcal{L}_{lat} = [\operatorname{MaxMean}({\bf X}_n,{\bf X}) - \operatorname{MaxMean}({\bf X}_p,{\bf X})]_{+}.
\end{equation}
Finally, our overall loss $\mathcal{L}_{lal}$ is given by $\mathcal{L}_{lal} = \mathcal{L}_{lac} + \mathcal{L}_{lat}$.
\vspace{-0.4em}
\subsection{Two-Stage Feature Retrieval}
\vspace{-0.2em}
An efficient feature retrieval pipeline is also critical for constructing a practical industrial-strength CSI system. Previous methods usually use an all-pairs strategy that includes computing the similarity between the query sample and each item in the database, which is time consuming. Moreover, there is a significant leap in complexity from the vector similarity measure ($\mathcal{O}(1)$) to the local alignment measure \textit{MaxMean} ($\mathcal{O}(n^2)$)~\cite{cai2016two}, which makes the all-pairs strategy even worse for ByteCover3. To solve this problem, we propose a two-stage pipeline with a hierarchical searching strategy for the retrieval of deep local embeddings. 

Given a query sample with $M$ local features, the first stage is to eliminate the database recordings that are highly unlikely to be a match. Specifically, for each local feature in the query, we search for its Top-$K$ nearest neighbors in the gallery of local features extracted in advance from all the database recordings, using the hierarchical navigable small world (HNSW) graphs \cite{malkov2018efficient}, with $K$ set to $50$. This results in a candidate set of $M\times K$ local matches for the given query, based on which our second stage of feature retrieval is further performed. Suppose that the $M \times K$ local matches originate from $D$ database recordings ($D \leq M\times K$ since some local matches may original from the same recordings), and thus our second stage is to compare the given query with each of the $D$ candidate recordings, based on the \emph{MaxMean} measure introduced above. The candidate recordings with the highest \emph{MaxMean} similarities are finally outputted as the retrieval results.

In practical use, the query is typically less than 60s, and thus we have $M \leq 5$ as our local features are extracted every 20s with overlap of 10 seconds. Therefore, in the second stage we only need to calculate the \emph{MaxMean} similarity for $M \times K \leq 250$ times, which is significantly less then the calculation needed in the all-pairs strategy.


\vspace{-1em}
\section{Experiments}
\label{sec:exp}

\vspace{-0.5em}

\subsection{Evaluation Settings and Training Details} 
\label{subsec:dataset}

\vspace{-0.2em}
The evaluation of ByteCover3 was conducted based on three public datasets: (1) \emph{SHS100K}~\cite{xu2018key}, which is collected from the \textit{Second Hand Songs} dataset, and consists of 8,858 cover groups and 108,523 recordings; (2) \emph{Covers80} \cite{ellis2007identifying}, which contains 160 recordings of 80 songs, with 2 covers per song; and (3) \emph{Da-TACOS} \cite{yesiler2020accurate}, which consists of 1000 cliques and 15,000~music performances. 

More specifically, the training subset of SHS100K was used to train ByteCover3, and to obtain short music clips for training, we randomly cut a segment from each training recording, where the segment duration is uniformly sampled between $6$s and $60$s. These short music clips were then mixed with the original full-track training samples to form the final training set. The testing of ByteCover3 was performed in a query-retrieval mode using the testing subset of SHS100K, Cover80 and Da-TACOS. For each query, we constructed a query set consisting of the original full-track recording, and $9$ music clips randomly cut from it, with the duration being $6$, $10$, $15$, $20$, $25$, $30$, $40$, $50$ and $60$ seconds respectively.

For the training of \ourname{}, the weights of the trained ByteCover2 model were used to initialize the ResNet-IBN module. Similar to ByteCover2, we implemented ByteCover3 in Pytorch framework and trained it using the Adam Optimizer. The learning rate and the batch size were set to $0.001$ and $128$ respectively. Every training batch contained synthetic short samples and full-length samples mixed in a $1:1$ ratio to improve the stability of training process.

During testing, the mean average precision (mAP) and the mean rank of the first correctly identified cover (MR1) were used as evaluation metrics. In our calculation of mAP and MR1, we set the similarity values to 0 for the database items that are blocked in the first retrieval stage. 

Moreover, since the queries were derived from snippets of some recordings in the result list, we ignored these recordings when calculating the evaluation metrics, to ensure that there is no leakage in the final detection performance.

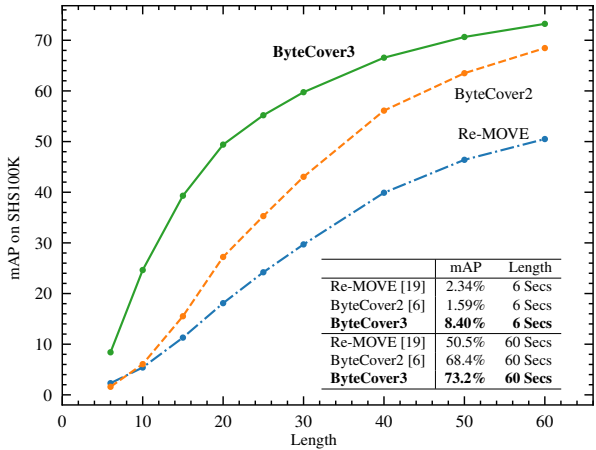
\begin{figure}[t]

	\resizebox{0.95 \columnwidth}{!}{

	\begin{tikzpicture}
        \node (image) at (0,0) {

            \includegraphics[width=.7\textwidth]{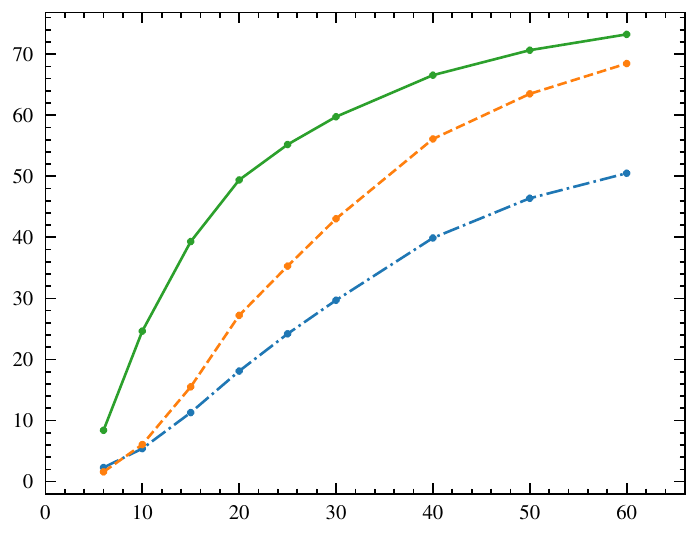}
            };
        \node[latex-] at (11em,7.5em){ByteCover2};
        \node[latex-] at (0em,10.1em){\textbf{\ourname}};
        \node[latex-] at (11em,5.2em){Re-MOVE};

        \node[latex-] at (0em,-13.7em){Length};
        \node[latex-] at (-18.4em,0em){\rotatebox{90}{mAP on SHS100K}};
        
        \node[latex-] at (7.8em,-6.5em) {\resizebox{0.6\columnwidth}{!}{
		\begin{tabular}[b]{l|cr}
		    \hline
			& mAP & Length  \\
			\hline

			Re-MOVE\cite{yesiler2020less} & 2.34\%  & 6 Secs \\
			ByteCover2\cite{du2022bytecover2} & 1.59\%  & 6 Secs \\
			\bf \ourname & \bf 8.40\%  & \bf 6 Secs \\
			\hline

			Re-MOVE\cite{yesiler2020less} & 50.5\% & 60 Secs  \\
			ByteCover2\cite{du2022bytecover2} & 68.4\%  & 60 Secs \\
			\bf \ourname & \bf 73.2\%  & \bf 60 Secs \\
			\hline
    	\end{tabular}}};
	
    \end{tikzpicture}
    }
	\hspace{-44mm}
        \vspace{-1em}
	\caption{Length of Queries vs. Performance.}
        \vspace{-1.5em}
	\label{fig:lengperf}
\end{figure}

\vspace{-0.6em}
\subsection{Comparison on Performance and Efficiency}
\label{ssec:comp}
\vspace{-0.5em}

Fig.~\ref{fig:lengperf} displays the mAP results of \ourname{} for different query lengths on the synthetic SHS100K test set, using Re-MOVE \cite{yesiler2020less} and ByteCover2 \cite{du2022bytecover2} as compared methods. As illustrated in the figure, our \ourname{} model achieves the best mAPs for all the query lengths. This clearly indicates the effectiveness of ByteCover3 for identifying short music queries. As for the task of identifying full-length recordings, our ByteCover3 is also competitive. As presented in Table~1 (the first three parts), the performances of ByteCover3 are comparable with the state-of-the-art method ByteCover2~\cite{du2022bytecover2} on all the three test sets in the full-length settings, even with an embedding size of $512$, which is smaller than those of ByteCover and ByteCover2.

To measure the effect of our proposed LAL loss on the performance of CSI, we conducted an ablation study on SHS100K-TEST for query length of 30s, by comparing \ourname{} with ByteCover2 and \emph{ByteCover2 + Local}. \emph{ByteCover2 + Local} modifies ByteCover2 with local features as done in \ourname{}, and it differs from \ourname{} in that it does not use the LAL loss. The lower part of Table~1 gives the comparison results. As shown in the table, \ourname{} achieves the highest mAP and lowest MR1 among the methods, which obviously proves the importance of LAL for short-query CSI.

Our last experiment is to test the time consumption of CSI, using the same setting as in \cite{du2022bytecover2}. As shown in Table~2, even equipped with local features, the retrieval speed of \ourname{} is still at the same scale with ByteCover2 using an embedding size of $128$. We owe this to the use of our two-stage feature retrieval pipeline. Please note that the inference time of ByteCover3 is two times longer than ByteCover2, because we split the input CQT spectrogram into chunks and the complexity of feature extraction is higher. However, the total time consumption of ByteCover3 is sitll similar with that of ByteCover2-128.

\begin{table}[!htbp]
  \small
  \centering
  \setlength\tabcolsep{12pt}
  \label{tab:com11}
\resizebox{ 0.95\columnwidth}{!}{
\begin{tabular}{c|ccc}
\hline
Model                                      & \#Dims. $\downarrow$ & mAP $\uparrow$ & MR1 $\downarrow$ \\ \hline
                                           & \multicolumn{3}{c}{Covers80 ~\cite{ellis2007identifying} (full)} \\ \hline
CQT-Net~\cite{yu2020learning}              & 300                  & 0.840          & 3.85             \\
ByteCover~\cite{du2021bytecover}           & 2048                 & 0.906          & 3.54             \\
ByteCover2 ~\cite{du2022bytecover2}        & 1536                 & \textbf{0.928} & \textbf{3.23}    \\

ByteCover3       & 512                 & \textbf{0.927} & \textbf{3.32}    \\ \hline

       & \multicolumn{3}{c}{\emph{Da-TACOS} \cite{yesiler2020accurate} (full)}        \\ \hline

ReMOVE~\cite{yesiler2020less}           & 2048                 & 0.525         & -            \\
ByteCover2 ~\cite{du2022bytecover2}        & 1536                 & \textbf{0.791} & \textbf{19.2}             \\
ByteCover3                                 & 512                  & 0.703         & 36.7    \\ \hline

                                           & \multicolumn{3}{c}{SHS100K-TEST ~\cite{xu2018key} (full)}        \\ \hline
CQT-Net~\cite{yu2020learning}              & 300                  & 0.655          & 54.9             \\
ByteCover~\cite{du2021bytecover}           & 2048                 & 0.836          & 47.3             \\
ByteCover2 ~\cite{du2022bytecover2}        & 1536                 & \textbf{0.864} & 39.0             \\
ByteCover3                                 & 512                  & 0.8242         & \textbf{37.0}    \\ \hline
                                           & \multicolumn{3}{c}{SHS100K-TEST ~\cite{xu2018key} (30~Secs)}         \\ \hline
ByteCover2 \cite{du2022bytecover2}         &  1536                 & 0.430         & 244.1               \\
ByteCover2 + local & 1536                 & 0.413          & 212.8         \\
ByteCover3                                 & 512                  & \textbf{0.734}          & \textbf{99.9}             \\ \hline

\end{tabular}
}

  \caption{Performance on different datasets and ablation study on 30 seconds version SHS100K-TEST.}
  \end{table}
\vspace{-1em}

\begin{table}[!htbp]
\centering
\small
\renewcommand\arraystretch{1.2}
  \setlength\tabcolsep{2pt}
\label{tab:time}
\resizebox{0.95\columnwidth}{!}{
\begin{tabular}{|c|c|c|c|c|} 
\hline
\multirow{2}{*}{\diagbox{Model}{Time (ms)}} & \multicolumn{2}{c|}{Embedding Extraction} & \multirow{2}{*}{Retrieval} & \multirow{2}{*}{Total}  \\ 
\cline{2-3}
                                             & Preprocess      & Inference         &                           &                         \\ 
\hline
Re-MOVE  \cite{yesiler2020less}                   & 5352 $\pm$ 123 & \textbf{60 $\pm$ 8.7}        & 360 $\pm$ 73                    &              5772           \\ 
\hline
ByteCover2-1536 \cite{du2022bytecover2}                               & \textbf{285 $\pm$ 31}  & 108 $\pm$ 15.2      & 2601 $\pm$ 384                  &                   2994      \\ 
\hline
ByteCover2-128 \cite{du2022bytecover2}                                   & 292 $\pm$ 39  & 105 $\pm$ 13.5      & \textbf{141 $\pm$ 13}                   &                  \textbf{538}       \\
\hline
\ourname{}-512                                & 290 $\pm$ 30  & 209 $\pm$ 14.6      & 237 $\pm$ 19                  &                 731      \\
\hline

\end{tabular}
}
\caption{Time consumption of different models in data preprocessing, model inference and retrieval phases respectively.}
\vspace{-1.2em}
\end{table}

\vspace{-1.0em}
\section{Conclusion}
\vspace{-0.5em}
In this paper, we propose to combine local feature matching and two-stage feature retrieval for efficient CSI of short music queries. A new loss termed LAL is designed to optimize the similarity measurement between songs with different length. Experimental results show that \ourname\ outperforms all benchmark models on three synthetic datasets for short-query CSI, while being highly efficient in local embedding extraction and hierarchical retrieval. For future work, we are currently studying to apply \ourname{} to other real-world applications such as set list identification, music matching with accurate timestamp and humming recognition.


\vfill\pagebreak

\bibliographystyle{IEEEbib}
\bibliography{6_refs}

\end{document}